\documentclass[conference]{IEEEtran}
\IEEEoverridecommandlockouts
\usepackage{cite}
\usepackage{algorithm}  
\usepackage{algpseudocode} 
\usepackage{colortbl}
\definecolor{lightgray}{gray}{0.7}
\usepackage{amsmath,amssymb,amsfonts}
\usepackage{bbm}
\usepackage{mathrsfs}
\usepackage{dutchcal}
\usepackage{graphicx}
\usepackage{textcomp}
\usepackage{xcolor}
\usepackage{threeparttable}
\usepackage{subfigure}
\allowdisplaybreaks
\def\BibTeX{{\rm B\kern-.05em{\sc i\kern-.025em b}\kern-.08em
    T\kern-.1667em\lower.7ex\hbox{E}\kern-.125emX}}
\begin{document}

\title{Mobility Aware Optimization in the Metaverse 
\\[-1.0ex]}
\author{\IEEEauthorblockN{Zhaohui Huang and Vasilis Friderikos}
\IEEEauthorblockA{Center of Telecommunication Research, 
King's College London,
London, U.K. \\
E-mail: \{zhaohui.huang, vasilis.friderikos\} @kcl.ac.uk}
\\[-8.0ex]
}

\maketitle

\begin{abstract}
Metaverse applications that incorporate Mobile Augmented Reality (MAR)  provide mixed and immersive experiences by amalgamating the virtual with the physical world. Notably, due to their multi-modality such applications are demanding in terms of energy consumption, computing and caching resources to efficiently support foreground interactions of participating users and rich background content. In this paper, the metaverse service is decomposed and anchored at suitable edge caching/computing nodes in 5G and beyond networks to enable efficient processing of background metaverse region models embedded with target AROs. To achieve that, a joint optimization problem is proposed, which explicitly considers the user physical mobility, service decomposition, and the balance between service delay, user perception quality and power consumption. A wide set of numerical investigations reveal that, the proposed scheme could provide optimal decision making and outperform other nominal baseline schemes which are oblivious of user mobility as well as do not consider service decomposition.   
\end{abstract}

\begin{IEEEkeywords}
Metaverse, 5G, Augmented Reality, Mobility
\end{IEEEkeywords}
\setlength{\parskip}{0em}

\section{Introduction}
\IEEEPARstart{I}{n} metaverse the digital and physical worlds are converging by utilizing mobile augmented/virtual reality technologies, edge computing, increased data rate support in 5G and beyond networks, digital twin and the proliferation of high end devices such as head-mounted displays \cite{xu2021wireless}\cite{dong2019deep}.
Users equipped with mobile augmented reality (MAR) devices can upload and analyze their environment through augmented reality customization to achieve appropriate augmented reality objects (AROs) and access the metaverse utilizing 5G mobile edge caching/computing enabled networks \cite{xu2022full}. In this emerging ecosystem,  rendering 3-dimentional (3D) AROs with the background virtual environment and updating it in the metaverse is demanding in terms of  energy consumption as well as in terms of caching and computing resources \cite{xu2022full}\cite{li2020rendering}. Hence, in addition to the fact that such applications require low latency they are also energy sensitive and face challenges in ensuring user quality of experience and providing reliable interactions within the metaverse \cite{xu2021wireless}\cite{xu2022full}. 

Generally speaking, a metaverse scene will in essence consisted by a background view as well as many objects in foreground interactions. The background view at a defined amalgamated virtual and physical location can be deemed as static or slowly changing \cite{xu2022full}\cite{guo2020adaptive}. A nominal background scene is the 3D model of the metaverse, a presentation of a related background virtual environment based on a certain user viewport \cite{guo2020adaptive}\cite{kato2021split}. Even though this background does not change frequently its size can reach tens of megabytes and the corresponding complexity of rendering related functionalities measured by computation load is also significant (e.g., 10 CPU cycles/bit) \cite{guo2020adaptive}\cite{yang2018communication}. On the other hand, objects related to foreground interactions  (such as for example user avatars) that are embedded in the metaverse scene change much more frequently. However, those are significantly less complex than the background scene (e.g., 4 CPU cycles/bit) \cite{li2020rendering}\cite{guo2020adaptive}. But, even though those objects are less complex their frequent changes require rendering them in a timely manner so that to avoid a considerable quality of experience degradation. Thus, in this paper, rendering for both foreground and background are deployed at at edge clouds (ECs) rather than only at terminals to make a full use of advanced caching and computing resources. Noticing that uploaded information are focused in foreground interactions while background content checking consumes not only computing resources but also lots of local cache to match and integrate AROs and related models of metaverse. The metaverse application could also be decomposed into computational and storage intensive functions which serve as a chain for better assignment and resource allocation \cite{huang2021proactive}. 

\begin{figure}[htb]
\centering
\includegraphics[width=0.75\linewidth]{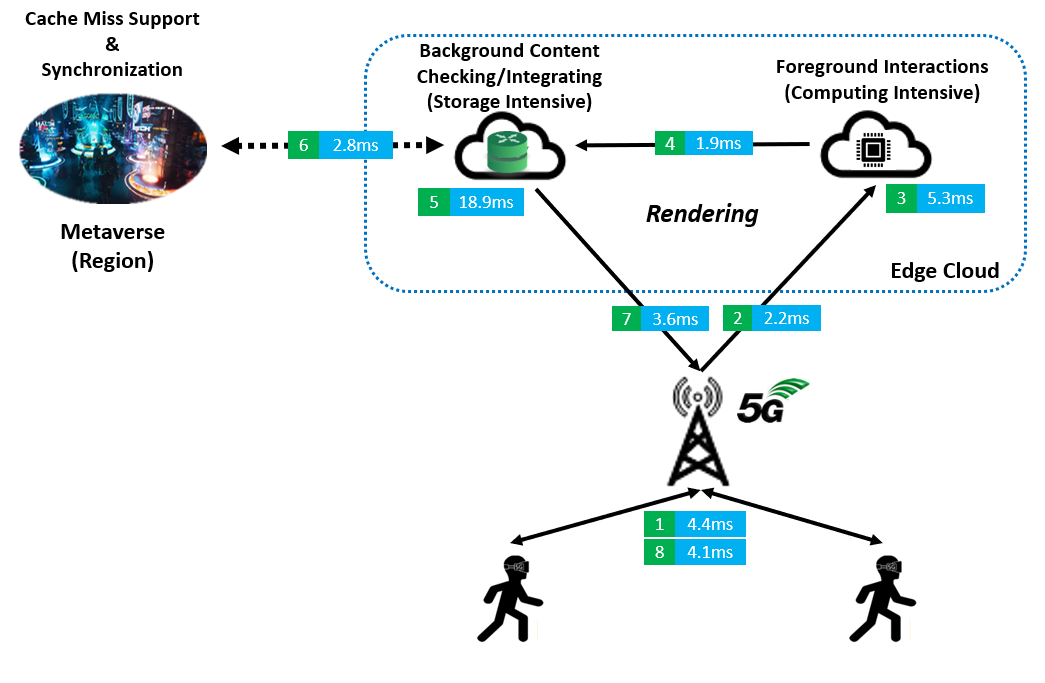}
\caption{The general work flow of a metaverse AR application.
}
\label{fig:workflow}
\end{figure}

The general work flow of a metaverse AR application supported by ECs is shown in Fig. \eqref{fig:workflow}. After triggered by certain behavior with foreground interactions \cite{zhang2021multi}, background contents like for example pre-cached 3D models and AROs are searched in the EC cache. If not found, the request is redirected to the metaverse region stored in a cloud deeper in the network. Finally, according to user's physical mobility and virtual orientation extracted from foreground interactions, the matched AROs and model are integrated into the frames and streamed to the user \cite{xu2022full}\cite{guo2020adaptive}. At the same time, updated information is sent to the metaverse region for synchronization so that the user could be aware of changes caused by other participants if they share the same metaverse region. Hereafter, we assume a nominal frame rate as 15 frames/second and the rendering happens at every other frame (\~ 133.2ms interval) \cite{cozzolino2022nimbus}\cite{niu2018learning}.

\begin{figure}
    \centering
    \includegraphics[width=0.75\linewidth]{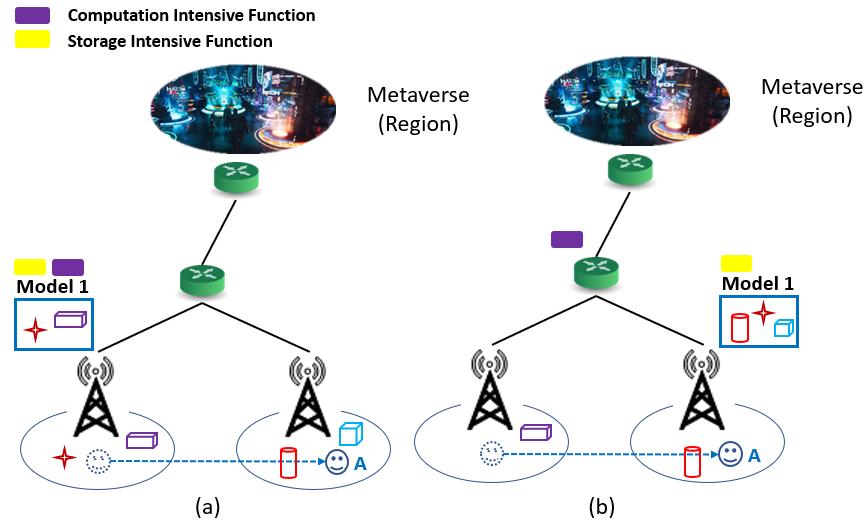}
    \caption{Illustrative toy example. Case (a) mobility is not considered and (b) physical mobility of the end user and metaverse service decomposition are considered with different renderings on pro-active resource allocation.}
    \label{fig: toyMetaAR}
\end{figure}

Fig. \eqref{fig: toyMetaAR} further reveals the difference between cases that consider the user mobility with service decomposition or not related to rendering requirements in the metaverse application. Clearly, when neglecting user mobility and service decomposition as shown by case (a), models, target AROs and metaverse functions are all cached close to the user's initial location. 
However, when user mobility and service decomposition is enabled, as shown in case (b), then service delivery becomes more flexible and efficient in terms of assigning requests and allocating network resources. Observe that in case (a), although user A is only one (wireless) hop away from the supporting EC, it requires 3 hops (one wireless and two wireline in the access network) after the mobility event. However, when mobility and EC resources are taken into consideration, the same user A in case (b) could enjoy a better maximum delay by facing 2 hops regardless of changing the point of attachment. Hence, in a high mobility scenario, it might not always be ideal to allocate requests and services as close as possible to the user's initial location. The AR contents in the model might be similar in terms of the viewport of different users. Hence, participating users should be aware of each other's updates and could share rendering functions to save on consumed resources. In this paper, we apply Structural Similarity (SSIM) proposed by \cite{wang2004image} for user perception experience. It is a widely accepted method that measures user perception quality of a image by comparing to its original version \cite{wang2004image}. Caching more models and AROs also causes more processing and transmission delay with energy consumption \cite{xu2021wireless}\cite{xu2022full}. Hence, the the joint optimization has to accept some potential loss due to constraints of computing and storage resources. In this paper, by considering explicitly the user mobility, service decomposition and models of metaverse regions with embedded AROs, the proposed optimization framework seeks a balance between user perception quality, power consumption and service delay. 

\section{System Model}
With the set $\mathbb{M}=\{1,2,...,M\}$ we express the available Edge Clouds (ECs) in the wireless network. With $r \in \mathbb{R}$, we denote the corresponding MAR service requests in the metaverse region that are generated by mobile users that are equipped with MAR devices (each user makes a single request). The starting location of the request $r$ is the access router where this user is initially connected to; this initial location is defined as $f(r)$. A user moves to a destination $k\in \mathbb{K}$ in the case of a mobility event (i.e., changing the point of attachment). In this paper, and without loss of generality, only adjacent access routers can be regarded as allowable destinations in the mobility event. A series of metaverse regions are set on ECs to interact with users. The corresponding metaverse region serving the user can be found through functions $A(f(r)), A(k)$. As explained earlier, each metaverse region is pre-deployed on an server close to the mobile network and its distance to an EC is also predefined. In this paper, As already mentioned in previous sections, a set of AROs is assumed to be embedded  across the  different non real time view streams. To this end, we first define a set $\mathbb{N}=\{1,2,...,N\}$ to represent the set of available AROs. The model available to each user has multiple rendering and we define them as a set $\mathbb{S}_r=\{1,2,...,S\}$. Then, the decision variable for proactively caching a model $s$ at the EC $j\in \mathbb{M}$ is denoted as $p_{sj}$. The subset $\mathbb{L}_{rs}$ represents the target AROs required by the user $r$ in related model $s\in \mathbb{S}_r$ and the size of each target ARO $l\in \mathbb{L}_{rs}$ is denoted as $O_l$. Finally, the decision variable for proactively caching an ARO required by a request $r$ is denoted as $h^s_{rl}$. More specifically, $p_{sj}$ and $h^s_{rl}$ can be written as follows,
\begin{equation}
p^{j}=\left\{
\begin{aligned}
1,&\;\text{if rendering the related model}\;s\;\text{at node}\;j, \\
0,&\;\text{otherwise}.
\end{aligned}
\right.
\end{equation}
\begin{equation}
h^s_{rl}=\left\{
\begin{aligned}
1,&\;\text{if ARO}\;l\;\text{required by request}\; r\;\text{embedded}\qquad\qquad\qquad\quad\\&\;\text{in the model}\;s\;\text{is cached}, \\
0,&\;\text{otherwise}.
\end{aligned}
\right.
\end{equation} 
In addition to the above, the following constraints should also be satisfied,
\begin{equation}
\begin{aligned}
\sum_{r \in \mathbf{R}}h^s_{rl} \leqslant 1,\; \forall j\in \mathbf{M},\;\forall s\in \mathbf{S_r},\; \forall l\in \mathbf{L_{rs}} \\
\end{aligned}
\label{cons_h1}
\end{equation}
\begin{equation}
\begin{aligned}
\sum_{s\in \mathbf{S_r}}\sum_{l \in \mathbf{L_{rs}}}h^s_{rl} \geqslant 1,\; \forall r\in \mathbf{R} \\
\end{aligned}
\label{cons_h2}
\end{equation} 
\begin{equation}
\begin{aligned}
\sum_{j \in \mathbf{M}}p_{sj} \geqslant h^s_{rl},\; \forall r\in \mathbf{R},\;\forall s\in \mathbf{S_r},\; \forall l\in  \mathbf{L_{rs}} \\
\end{aligned}
\label{cons_h3}
\end{equation}
\begin{equation}
\begin{aligned}
h^s_{rl} \leqslant h^s_{rl} \sum_{j \in \mathbf{M}}p_{sj},\; \forall r\in \mathbf{R},\;\forall s\in \mathbf{S_r},\; \forall l\in  \mathbf{L_{rs}} \\
\end{aligned}
\label{cons_h4}
\end{equation}
Constraints in \eqref{cons_h1} ensure that each ARO can be cached at most once in a related model. Constraints in \eqref{cons_h2} ensure that at least one model and an ARO is required to compose a valid request. Constraints in \eqref{cons_h3} guarantee that the ARO must be allocated to a model first before deciding to proactively cache it and constraints in \eqref{cons_h4} certify that any ARO planned to be stored in this model should not be cached when deciding not to proactively cache the model at all. Constraints \eqref{cons_h3} and \eqref{cons_h4} together ensure that either the model or ARO cannot be handled alone during the formulation.

Denote $B_j$ as the bandwidth allocated to the user's resource block and $\gamma_{rj}$ as the Signal to Interference plus Noise Ratio (SINR) of the user $r$ at the node $j$. Denote $P^{tran}_{rj}$ as the current transmit power of the user $r$ at the node $j$, $P_i$ as the base station power, $H_{rj}$ as the channel gain, $N_j$ as the noise and $a$ as the path loss exponent and $d_{rj}$ as the Euclidean Distance between the user and the access router in the cell. Furthermore, a nominal Rayleigh fading channel is utilized to model the channel between a 5G access point and the users\cite{gemici2021modeling}. In this case, the channel gain $H_{rj}$ can be written as $H_{rj}=\sqrt{\frac{1}{2}}(t+t^{'}J)$, where $J^2=-1$, $t$ and $t^{'}$ are random numbers following the standard normal distribution \cite{cho2010mimo}. Then, the SINR $\gamma_{rj}$ in a 5G wireless network can be written as follows \cite{cho2010mimo},
\begin{equation}
\begin{aligned}
\gamma_{rj}=\frac{P^{tran}_{rj}H^2_{rj}d^{-a}_{rj}}{N_j+\sum_{i\in \mathbf{M}, i\neq j}P_iH^2_{ri}d^{-a}_{ri} }
\end{aligned}
\end{equation}
Denote the data rate as $g\in\mathbb{G}$ and the decision variable $e_{rg}$ to decide whether to select the data rate $g$ for user $r$,
\begin{equation}
e^{rg}=\left\{
\begin{aligned}
1,&\;\text{if data rate}\;g\;\text{is selected}\;\text{for user}\;r, \\
0,&\;\text{otherwise}.
\end{aligned}
\right.
\end{equation}
Noticing that the chosen data rate can also be written in Shannon Formula as $B_j \log_2 (1+\gamma_{rj})$. After choosing a data rate as $ge_{rg}$ for the user, the transmit power $P^{tran}_{rj}$ can be written as follows,
\begin{equation}
\begin{aligned}
P^{tran}_{rj}=\frac{N_j+\sum_{i\in \mathbf{M}, i\neq j}P_iH^2_{ri}d^{-a}_{ri}}{H^2_{rj}d^{-a}_{rj}}(2^{\frac{ge_{rg}}{B_j}}-1)
\end{aligned}
\label{transpower}
\end{equation}
Noticing that $2^{\frac{ge_{rg}}{B_j}}=(1-e_{rg})+e_{rg}2^{\frac{g}{B_j}}$ and should satisfy the following constraint to ensure that a user could only select one data rate,
\begin{equation}
\begin{aligned}
\sum_{g\in\mathbb{G}}e_{rg}=1, \forall r\in\mathbb{R}
\end{aligned}
\label{cons_h7}
\end{equation}

In a similar manner with \cite{huang2021proactive}, computational intensive and storage intensive MAR functionalities are defined as $\eta$ and $\varrho$, respectively. In addition, the corresponding execution location for a functionality is denoted as $x_{ri}$ and $y_{ri}$ respectively \cite{huang2021proactive}. In a mobility event, $u_{f(r)k} \in [0,1]$ is defined to represent the probability of a user moving from the initial location to an allowable destination, where adjacent servers $\{f(r),k\} \subset \mathbb{M}$. Such probabilities can also be learned from historical data, which are readily available from mobile operators. The size of foreground interactions is denoted as $F^{fore}_{\eta r}$, the size of pointers inside used for matching AROs is denoted as $F_{\varrho r}$ and the size of the related model $s$ used for background content checking is $F^{back}_{sr}$ \cite{huang2021proactive}\cite{guo2020adaptive}. During the matching and background content checking process, the target AROs or background content are possibly not pre-cached in the local cache and such case is known as a "cache miss" (otherwise there is a "cache hit"). Whenever confronted with a cache miss, the request is redirected to the metaverse region stored in a core cloud deeper in the network and suffers from an extra latency $D$ as penalty. After rendering, the model and target AROs are integrated into a compressed final frame for transmission and its compressed size is denoted as $F^{res}_{sr}$.

At first, a joint optimization scheme considering the balance between the service delay, the perception quality and the power consumption is designed to track the Metaverse AR requests in the EC supported network. The cache hit/miss is captured by the decision variable $z_{rj}$ and can be written as follows,
\begin{equation}
z_{rj}=\left\{
\begin{aligned}
1 ,&\;\text{if} \sum_{l \in \mathbf{L_{rs}}}\sum_{s \in \mathbf{S_r}}p_{sj}h^s_{rl} \geqslant L_{rs}, \\
0 ,&\;\text{otherwise}.
\end{aligned}
\right.
\end{equation}

In \cite{huang2021proactive}, constraints (10b) and (10d) reveal the cache limitation and the cache hit/miss relation. In this paper, they should be rewritten as follows,
\begin{equation}
\begin{aligned}
&\sum_{r \in \mathbf{R}}\sum_{l \in \mathbf{L_{rs}}}\sum_{s \in \mathbf{S_r}} p_{sj}h^s_{rl}O_{l}\leq \Theta_j,\forall j\in \mathbf{M}\\
\end{aligned}
\label{cons_h5}
\end{equation}
\begin{equation}
\begin{aligned}
&\sum_{l  \in \mathbf{N}}\sum_{s \in \mathbf{S_r}} h^s_{rl} + \epsilon \leq L_{rs} + U(1-q_{rj}) \: \forall j\in \mathbf{M},  r \in \mathbf{R}\\
\end{aligned}
\label{cons_h6}
\end{equation}
where $\Theta_j$ is the cache capacity at node $j$. In \eqref{cons_h6}, to rewrite the either-or constraint that $\sum_{l  \in \mathbf{N}}\sum_{s \in \mathbf{S_r}} h^s_{rl} < L_{rs} $ or $z_{rj}=1$, we define $\epsilon$ as a small tolerance value, $U$ as a large arbitrary number and $q_{rj}$ as a new decision variable satisfying $1-q_{rj}=z_{rj}$ \cite{huang2021proactive}. Requiring more pre-cached models and embedded AROs in a request naturally lead to an extra burden for the matching function. More specifically, the processing delay of the computational intensive function can be written as,
\begin{equation}
\begin{aligned}
V_{rj}=\frac{\omega_{\eta} F^{fore}_{\eta r}}{f_V^j}
\end{aligned}
\end{equation}
Similarly, the processing delay of the matching and background content checking function can be written as,
\begin{equation}
\begin{aligned}
W_{rj}=\frac{\omega_{\varrho} (F_{\varrho r}+\sum_{l \in \mathbf{L_{rs}}}\sum_{s \in \mathbf{S_r}}p_{sj}h^s_{rl}O_l+\sum_{s \in \mathbf{S_r}}F^{back}_{sr}p_{sj})}{f_V^j}
\end{aligned}
\end{equation}
where $\omega_{\eta}$ and $\omega_{\varrho}$ (cycles/bit) represent the computation load of foreground interaction and background matching, $f^j_V$ is the virtual CPU frequency (cycles/sec), $F_{\varrho r}$ are the size of uploaded pointers of AROs in foreground interactions \cite{huang2021proactive}\cite{guo2020adaptive}. 
When finding the target AROs during matching, their pointers included by foreground interactions should also be transferred to the metaverse for updating. Finally, the final frame integrating the model and target AROs of is transmitted back to the user. Hence, the wired transmission delay for each user after processed by functions can be written as follows,
\begin{equation}
\begin{aligned}
\sum_{s \in \mathbf{S_r}}\sum_{j \in \mathbf{M}}(C_{jA(f(r))}+C_{jA(k)})p_{sj}+\\
(C_{A(f(r))f(r)}+ \sum_{k \in \mathbf{K}}C_{A(k)k}u_{f(r)k})
\end{aligned}
\end{equation}
where $u_{f(r)k}$ is the moving probability from the initial location $f(r)$ to a potential destination $k$. In previous expressions, the product of decision variables $p_{sj}h^s_{rl}$ and $p_{sj}y_{rj}$ exists and creates a non-linearity. In addition, note that when executing the matching function at the location $j$ ($W_{rj}y_{rj}$), the product of decision variables $p_{sj}h^s_{rl}y_{rj}$ also appears. To linearize the expressions above, so that to utilize linear integer programming solution methodologies, new auxiliary decision variables are brought in. A new decision variable $\alpha_{rsj}$ is introduced as $\alpha_{rsj}=p_{sj}y_{rj}$ and the constraints should be added as follows,
\begin{equation}
\begin{aligned}
 \alpha_{rsj} \leqslant p_{sj}, \; 
 \alpha_{rsj} \leqslant y_{rj},\;
 \alpha_{rsj} \geqslant p_{sj}+y_{rj}-1 \label{a1}\\
\end{aligned} 
\end{equation}

Similarly, a new decision variable $\beta_{rslj}$ is introduced as $\beta_{rslj}=p_{sj}h^s_{rl}$ and the constraints should be added as follows, 
\begin{equation}
\begin{aligned}
 \beta_{rslj} \leqslant p_{sj},\; 
 \beta_{rslj} \leqslant h^s_{rl} ,\; 
 \beta_{rslj} \geqslant p_{sj}+h^s_{rl}-1 \label{a2}\\
\end{aligned}
\end{equation}


Also, note that $p_{sj}$ is a binary decision variable and causes $p_{sj}=p^2_{sj}$. Thus, we have $p_{sj}h^s_{rl}y_{rj}=\alpha_{rsj}\beta_{rslj}$. A new decision variable $\lambda_{rslj}$ is introduced as $\lambda_{rslj}=\alpha_{rsj}\beta_{rslj}$ and the following constraints should be added as follows, 
\begin{equation}
\begin{aligned}
\lambda_{rslj} \leqslant \alpha_{rsj},\; \lambda_{rslj} \leqslant \beta_{rslj},\;\lambda_{rslj} \geqslant \alpha_{rsj}+\beta_{rslj}-1 \label{a3}\\
\end{aligned}
\end{equation}

Hence, the product $W_{rj}y_{rj}$ can be rewritten as follows,
\begin{equation}
\begin{aligned}
\frac{\omega_{\varrho} (F_{\varrho r}y_{rj}+\sum_{l \in \mathbf{L_{rs}}}\sum_{s \in \mathbf{S_r}}\lambda_{rslj}O_l+\sum_{s \in \mathbf{S_r}}F^{back}_{sr}\alpha_{rsj})}{f_V^j}
\label{storage_intensive_comp}
\end{aligned}
\end{equation}

By checking whether users share the same metaverse region by $A(f(t))=A(f(r)), \{t,r\}\subset\mathbb{R}$, we can ensure the user could also view other updates happening in the same metaverse region. Based on the previous modelling of wireless channel, the wireless transmission delay in a mobility event can be written as follows,
\begin{equation}
\begin{aligned}
&\sum_{r\in \mathbf{R}}\frac{F^{fore}_{\eta r}+\sum_{t\in\mathbf{R}, A(f(t))=A(f(r))}\sum_{s\in\mathbf{S_r}}p_{sj}F^{res}_{st}}{ge_{rg}}+\\&\sum_{r\in \mathbf{R}}\sum_{k\in\mathbf{K}}u_{f(r)k}\frac{F^{fore}_{\eta r}+\sum_{t\in\mathbf{R}, A(t)=A(k)}\sum_{s\in\mathbf{S_r}}p_{sj}F^{res}_{st}}{ge_{rg}}
\label{wireless}
\end{aligned}
\end{equation}

Noticing that with the aforementioned definition of $e_{rg}$ and related constraint \eqref{cons_h7}, $\frac{1}{e_{rg}}$ can be replaced by $e_{rg}$ for linearization. By introducing a new decision variable $\phi_{rlsg}$ with following constraints,
\begin{equation}
\begin{aligned}
\phi_{rsg} \leqslant e_{rg},\;\phi_{rsg} \leqslant p_{sj},\;&\phi_{rsg} \geqslant e_{rg}+p_{sj}-1 \label{a4}\\
\end{aligned}
\end{equation}

Thus, the previous formula \eqref{wireless} can be updated as follows,
\begin{equation}
\begin{aligned}
&\frac{1}{g}\sum_{r\in \mathbf{R}}(1+\sum_{k\in\mathbf{K}}u_{f(r)k}) (F^{fore}_{\eta r}e_{rg}+\\&\sum_{t\in\mathbf{R}, A(f(t))=A(f(r))}\sum_{s\in\mathbf{S_r}}\phi_{rsg}F^{res}_{st} )
\label{wireless_new}
\end{aligned}
\end{equation}

Based on the above derivations and inline with \cite{huang2021proactive}, the overall latency can be written as follows,
\begin{equation}
\begin{aligned}
L=&\eqref{wireless_new}+\sum_{r \in \mathbf{R}}\sum_{i \in \mathbf{M}}(C_{f(r) i}+V_{ri})x_{ri}+\\
&\sum_{r \in \mathbf{R}}\sum_{i \in \mathbf{M}}\sum_{j \in \mathbf{M}} (\eqref{storage_intensive_comp}+ C_{ij}\xi_{rij}+C_{A(f(r))f(r)}+\psi_{rj}D  )+\\ \\
&\sum_{s \in \mathbf{S_r}}\sum_{j \in \mathbf{M}}(C_{jA(f(r))}+C_{jA(k)})p_{sj}+ \\
&\sum_{r \in \mathbf{R}}\sum_{k \in \mathbf{K}}(C_{A(k)k}+C_{ki}x_{ri})u_{f(r)k}
\end{aligned}
\end{equation}
where $V_{ri}$ is the processing delay of computational intensive function \cite{huang2021proactive}. Note that $L_{max}$ here denotes the maximum allowed service latency and therefore the following holds, $\frac{L}{L_{max}} \in [0,1]$.

The energy cost of the system during each service time slot is measured by its total consumed power. The total power consists of the transmission power and the CPU processing power at target ECs. We denote the required CPU processing power of the user $r$ at the node $j$ as $P^{cpu}_{rj}$ and the CPU chip architecture coefficient as $k_0$ (e.g. $10^{-15}$) \cite{dong2019deep}. Then, the power per CPU cycle at the EC can be achieved through $k_0(f^j_V)^2$ (Watt/cycle) based on measurements in \cite{zhang2013energy}\cite{miettinen2010energy}. Thus, the total power consumption can be written as follows,
\begin{equation}
\begin{aligned}
P &=\sum_{r\in\mathbf{R}}\sum_{j\in\mathbf{M}}(P^{tran}_{rj}+P^{cpu}_{rj})\\
&=\sum_{r\in\mathbf{R}}\sum_{j\in\mathbf{M}}(\frac{N_j+\sum_{i\in \mathbf{M}, i\neq j}P_iH^2_{ri}d^{-a}_{ri}}{H^2_{rj}d^{-a}_{rj}}(2^{\frac{ge_{rg}}{B_j}}-1)+\\
&\quad\qquad\qquad k_0(f^j_V)^2 (W_{rj}y_{rj}+V_{rj}x_{rj})f^j_V  )
\end{aligned}
\end{equation} 
$P_{max}$ represents the maximum allowable total power of the system. It also has $\frac{P}{P_{max}} \in [0,1]$.

SSIM is applied to reveal the quality of perception experience. In this paper, the video coding scheme (e.g. H.264) and frame resolution (e.g. 1280$\times$720) are assumed as pre-defined \cite{kato2021split}. Then SSIM is mainly affected by data rate and a concave function could be applied to reveal the relation between them\cite{kato2021split}. Hence, the SSIM value is denoted as $c$ and consists a set for each ARO under corresponding data rate as $\mathbb{SSIM}_{l}, l\in\mathbb{L}_r$. The overall quality of perception experience $Q$ can be written as follows,
\begin{equation}
\begin{aligned}
Q=\sum_{r\in\mathbf{R}}\sum_{l\in\mathbf{L_r}}\sum_{g\in\mathbf{G}}\sum_{c\in\mathbf{SSIM}_{l}} e_{rg}c 
\end{aligned}
\end{equation}
Similarly, $Q_{max}$ here denotes the maximum available quality of perception experience and satisfies $\frac{Q}{Q_{max}} \in [0,1]$.

We denote the weight parameter is denoted as $\mu \in [0,1]$ and the joint optimization problem can eventually be written as follows,
\begin{subequations}
\begin{align}
\mathop{min}
&\; \frac{\mu}{2}(\frac{L}{L_{max}}-\frac{Q}{Q_{max}})+(1-\mu) \frac{P}{P_{max}}
\label{JOP:1}
\\
\nonumber
\\
\text{s.t.}\;& z_{rj} = 1-q_{rj}, \: \forall j\in \mathbf{M},  r \in \mathbf{R}
\label{JOPcon:1}
\\
&\sum_{r \in \mathbf{R}} (x_{rj}+y_{rj})\leq \Delta_j,\forall j\in \mathbf{M}
\label{JOPcon:2}
\\
&\sum_{j  \in \mathbf{M}} x_{rj}=1,\forall r\in \mathbf{R}
\label{JOPcon:3}
\\
&\sum_{j  \in \mathbf{M}} y_{rj} =1,\forall r \in \mathbf{R}
\label{JOPcon:4}
\\
&\xi_{rij} \leq x_{ri}, \: \forall r \in \mathbf{R}, i,j \in  \mathbf{M}
\label{JOPcon:5}
\\
&\xi_{rij} \leq y_{rj}, \: \forall r \in \mathbf{R}, i,j \in  \mathbf{M}
\\
&\xi_{rij} \geq x_{ri} + y_{rj} -1, \: \forall r \in \mathbf{R}, i,j \in  \mathbf{M}
\\
&\psi_{rj} \leq z_{rj}, \: \forall r \in \mathbf{R}, j \in  \mathbf{M}
\\
&\psi_{rj} \leq y_{rj}, \: \forall r \in \mathbf{R}, j \in  \mathbf{M}
\\
&\psi_{rj} \geq z_{rj} + y_{rj} -1, \: \forall r \in \mathbf{R}, j \in  \mathbf{M}
\label{JOPcon:10}
\\
&x_{rj}, y_{rj},p_{sj}, h^s_{rl}, z_{rj}, q_j\in  \{0,1\},\nonumber\\
&\alpha_{rsj},\beta_{rslj},\lambda_{rslj}, \phi_{rslg},\psi_{rj}, \xi_{rij}\in \{0,1\}, \nonumber\\
&\forall r \in \mathbf{R},  j\in \mathbf{M},  l \in \mathbf{L_{rs}}, s\in \mathbf{S_r}
\label{JOPcon:11}
\\
&\eqref{cons_h1},\;\eqref{cons_h2},\;\eqref{cons_h3},\;\eqref{cons_h4},\;\eqref{cons_h7},\;\eqref{cons_h5},\;\eqref{cons_h6},\;\eqref{a1},\;\eqref{a2},\;\eqref{a3},\;\eqref{a4}\nonumber
\end{align}
\end{subequations}

As mentioned earlier, the constraint \eqref{JOPcon:1} together with constraints \eqref{cons_h1} to \eqref{cons_h4} reveal the relation between pre-caching decisions and the cache miss/hit for each request \cite{huang2021proactive}. The constraint \eqref{JOPcon:2} is the virtual machine limitation while \eqref{JOPcon:3} and\eqref{JOPcon:4} guarantee the once execution of each function of a request at a single server as explained in \cite{huang2021proactive}. The constraints \eqref{a1} to \eqref{a3} and \eqref{JOPcon:5} to \eqref{JOPcon:10} are auxiliary and required to solve the product of decision variables for linearization.

\section{Numerical Investigations}
In this section, the effectiveness of the proposed optimization scheme, which will be referred to as Optim in the sequel, is investigated 
and compared with a number of nominal/baseline schemes.
Same as in \cite{huang2021proactive}, a nominal tree-like network topology is applied with 20 ECs in total and 6 ECs being activated for the current metaverse AR service and 30 requests are sent by MAR devices. The remaining available resources allocated for metaverse AR support within an EC are assumed to be CPUs with frequency from 4 to 8 GHz, 4 to 8 cores and $[100,400]$MBytes of cache memory \cite{dong2019deep}\cite{huang2021proactive}. Each request requires a single free unit for each service function, such as for example a Virtual Machine (VM) \cite{liu2018edge}. Up to 14 available VMs are assumed in each EC, with equal splitting of the available CPU resources \cite{huang2021proactive}. Note that different view ports lead to different models of the metaverse \cite{guo2020adaptive}. Hereafter, up to 4 different models per user can be cached and are similar for each user. All target AROs must be integrated with the corresponding model and rendered within the frame before being streamed to the end user based on a matched result. 
The size of pointers used for matching are only a few bytes in size and hence their transmission and processing latency are neglected in the simulations. The set of available data rates is $\{2,3,...,8\}$Mbps and its corresponding SSIM values set is $\{0.955, 0.968,...,0.991\}$ \cite{kato2021split}. The carrier frequency of a nominal 5G access point is set to 2GHz, its transmit power is assumed to be 20dBm with the maximum of 100 resource blocks and we assume, without loss of generality, that each user can utilize one resource block\cite{korrai2019slicing}\cite{li2020computing}. The noise power is $10^{-11}$W and the path loss exponent is 4\cite{korrai2019slicing}. The location of the users is  randomly generated as well as the potential destinations. Furthermore, we assume that each cell has a radius as 250m in the 5G wireless network \cite{korrai2019slicing}. As mentioned earlier, we assume a predefined video coding scheme, namely the H.264 with a fixed frame resolution as 1280$\times$720 \cite{kato2021split} in RGB (8 bits per pixel). Based on the given resolution, the size of foreground interactions after decoding and compressing can be calculated by multiplying the coefficients $\frac{5}{9}$ and $10^{-3}$ \cite{guo2020adaptive}. 

Three baseline schemes implementing edge  caching decisions are also implemented for comparison. These are the Random Selection Scheme (RandS), the Closest First Scheme (CFS) \cite{tocze2019orch} and the utilization based scheme (UTIL) \cite{sonmez2019fuzzy}. The RandS scheme operates a random EC selection, while the other two schemes both select the closest EC to the user's initial location. The CFS scheme also accepts the second closest one as a back up choice whilst the UTIL scheme set a capacity boundary to ensure an ideal EC working state; we assume that this is 80\% occupied of available VMs  \cite{tocze2019orch}\cite{sonmez2019fuzzy}.

\begin{figure}[htb]
\centering
\includegraphics[width=0.7\linewidth]{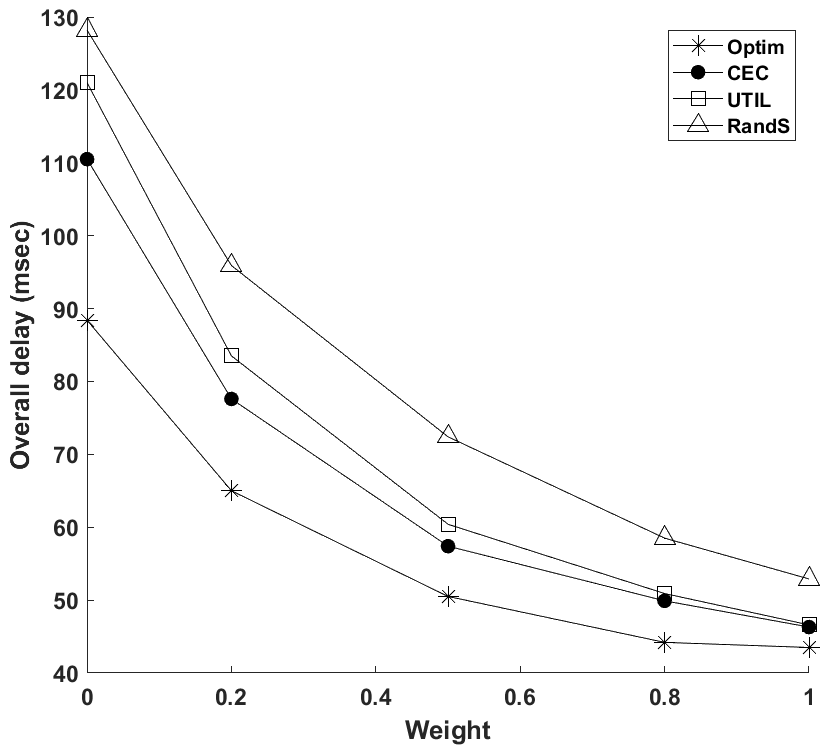}
\caption{Overall Delay with Weight $\mu$ (6 EC, 30 Requests, EC Capacity is 14 and total mobility probability is 1)}
\label{fig:W&Delay}
\end{figure}

\begin{figure}[htb]
\centering
\includegraphics[width=0.7\linewidth]{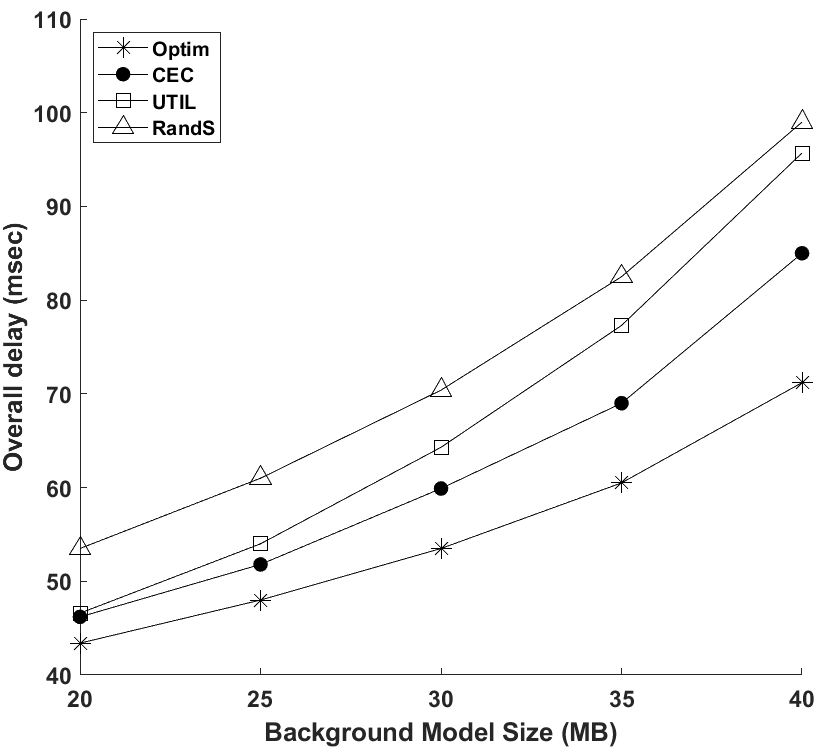}
\caption{Overall Delay with Background Model Size (6 EC, 30 Requests, EC Capacity is 14 and total mobility probability is 1)}
\label{fig:Back&Delay}
\end{figure}

According to Fig. \eqref{fig:W&Delay}, the service delay for each request drops, as expected, with an increasing weight $\mu$. With a larger weight, the Optim scheme tends to select a larger data rate and proactively cache fewer AROs, which lead to a smaller overall delay. Compared to the CFS scheme, for example, the gain in delay of the Optim scheme ranges from 6.4\% to 24.8\%.  As shown by Fig. Table \eqref{tab:P&SSIM}, the larger data rate caused by the increasing weight also leads to an increased perception quality for each target ARO. However, the power consumption in this case shows an ascending trend as revealed in Table \eqref{tab:P&SSIM}. Thus, a trade-off exists and power consumption should be considered vis-a-vis with the gains it offers  in delay and quality. By selecting a suitable weight, a balance could be achieved via the proposed Optim scheme between delay, quality and power consumption. As eluded before, the UTIL scheme has a stricter EC capacity limit and hence it is the most sensitive scheme to the weight. Fig. \eqref{fig:Back&Delay} further reveals how delay is impacted with an increase to the background model size. When the EC has sufficient resources the delay increase proportionally to the background model size. Hence, the delay of the Optim scheme is at first 6.3\% better than the CFS scheme and it increases by 2.5\% for each extra MB in background model size. But, when the EC becomes congested, the proposed Optim scheme has to accept sub-optimal solutions. However, the baseline schemes are much easier to excess the resource limit and suffer from a penalty. The gap between the CFS scheme and the Optim scheme could even reach 18.5\%. Thus, they become much worse in delay and more sensitive to the background model size. 

\begin{table}[htb]
\caption{Average Perception Quality SSIM and Power Consumption with Weight $\mu$ }
    \centering
    \begin{tabular}{c|c|c|c|c|c|c|c}
    \hline
         \textbf{Weight $\mu$}&0&0.2&0.5&0.8&1\\
         \hline
         \textbf{SSIM}&0.958&0.979&0.986&0.990&0.992\\
         \hline
         \textbf{Power (W)}&0.88&1.09&1.32&1.41&1.45\\
         \hline
         
    \end{tabular}
    \label{tab:P&SSIM}
\end{table}

\begin{table}[htb]
\caption{Overall Delay in no mobility event \\($\mu=1$, 6 ECs, 30 requests and EC Capacity is 14)}
    \centering
    \begin{tabular}{c|c|c|c|c|c|c}
    \hline
         \textbf{Scheme}&\textbf{Optim}&\textbf{CFS}&\textbf{UTIL}&\textbf{RandS}\\
         \hline
         \textbf{Delay (ms)}&37.7&38.6&38.7&44.3\\
         \hline
    \end{tabular}
    \label{tab:no_mob}
\end{table}
Observe that according to Table \eqref{tab:no_mob}, even in the case where there is no mobility (i.e., users are stationary), the proposed Optim scheme still manages to outperform other baseline schemes because its service decomposition could better avoid potential EC overloading. Therefore, the proposed Optim scheme has an obvious advantage over baseline schemes especially during network congestion episodes and a high user physical mobility environment.

\section{Conclusions}
Experiences in the metaverse  are expected to be significantly demanding in terms of energy consumption, persistent high data rate support and advanced edge caching/computing capabilities in 5G and beyond networks.
In this paper, a joint optimization framework is proposed that explicitly considers the model rendering, user mobility and service decomposition to achieve a balance between power consumption, user perception quality and service delay for content rich metaverse type of applications. Compared to nominal schemes which are mobility oblivious, the proposed optimization framework is able to provide an average reduction of 11.8\% to 35.6\% in terms of delay without sacrificing quality of experience or increasing the required power consumption.  

\bibliographystyle{IEEEtran}
\bibliography{bibliography.bib}

\end{document}